\documentclass[preprint,aps]{revtex4}

\usepackage{graphicx}% Include figure files
\usepackage{dcolumn}% Align table columns on decimal point
\usepackage{bm}% bold math
\usepackage{epstopdf}
\begin{document}

\title{Explosive growth in biased dynamic percolation on two-dimensional regular lattice networks}

\author{Robert M. Ziff}
\email{rziff@engin.umich.edu}
 \affiliation{Center for the Study of Complex Systems and Dept.\ of Chemical Engineering \\
University of Michigan, Ann Arbor MI 48109-2136 USA.}

\date{\today}

\begin{abstract}
The growth of two-dimensional lattice bond percolation clusters through a cooperative Achlioptas-type of process, where the choice of which bond to occupy next depends upon the masses of the clusters it connects, is shown to go through an explosive, first-order kinetic phase transition with a sharp jump in the mass of the largest cluster as the number of bonds is increased.  The critical behavior of this growth model is shown to be of a different universality class than standard percolation.   

\end{abstract}

\maketitle

%\section{\label{sec:Intro} Introduction}

Percolation concerns the formation of long range connectivity in systems \cite{StaufferAharony94}, and has applications to numerous practical problems including conductivity in composite materials, flow through porous media, and polymerization \cite{Sahimi94}.  In the usual random (Bernouilli) percolation, bonds are formed randomly and independently throughout the system, and at a critical concentration $p_c$ of occupied bonds, a finite fraction of the sites (vertices) are connected together and percolation takes place through a continuous or second-order transition.  
By the universality hypothesis, all large-scale properties near the transition point (i.e., the fractal dimension, critical exponents) are the same in a given dimensionality, irrespective of the microscopic details of the model \cite{StaufferAharony94}.

Recently, Achlioptas et al.\ \cite{AchlioptasDSouzaSpencer09} modified the growth of percolation clusters to produce a first-order kinetic transition on the mean-field-like random graph, through a procedure that is known as an Achlioptas process.  
In this optimization process, introduced to study problems in graph theory, two alternate choices of adding a bond are considered, and a specific strategy to favor a desired result (such as delaying or accelerating the appearance of a giant component, or the formation of a Hamiltonian cycle) is followed \cite{BeveridgeBohmanFriezePikhurko07,KrivelevichLohSudakov09}.    In Ref.\ \cite{AchlioptasDSouzaSpencer09}, the authors considered a process of growth of clusters in the Erd\H os-R\'enyi random graph network model  in which two unoccupied edges between clusters are chosen at random, and the one that minimizes the product of the two connecting cluster masses is preferentially chosen as the next occupied bond.   This process led to a sharp jump in the growth dynamics, which the authors labeled as explosive percolative growth, in contrast to the normal percolation phase transition, on both regular and random lattices, where the transition is second-order and continuous.

Erd\H os-R\' enyi random networks are formed by randomly linking pairs of points, irrespective of their distance apart.  These networks are essentially mean-field and infinite-dimensional, and for this problem, only the mass of the clusters needs to be kept track of.  There is a close connection between the percolation on Erd\H os-R\'enyi networks and percolation on the infinite-dimensional Bethe lattice and polymerization of non-looping branched polymers, whose percolation or gelation theory goes back to Flory \cite{Flory41}.  When bonds are added randomly between sites in the random graph, the net effective probability that a cluster of mass $s_1$ and a cluster of mass $s_2$ are joined is proportional to the product of their masses $s_1 s_2$.  \ This leads to the product kernel of the polymer growth process as described by the Smoluchowski equation, and this growth process is equivalent to percolation on the Bethe lattice \cite{Ziff80}.  Choosing the bond with the minimum product leads to the minority product rule (PR) of Ref.\ \cite{AchlioptasDSouzaSpencer09}.

While random and scale-free networks have recently received a great deal of attention \cite{NewmanBarabasiWatts06}, many actual networks of interaction, whether physical or social, are restricted spatially and bonds can only form between close neighbors.  The two-dimensional percolation model is the strongest form of such a restriction and has been the subject in intense study for over fifty years \cite{StaufferAharony94}.  Random percolation can be viewed as the result of a growth process where bonds are added to a single cluster (as in the Leath algorithm) or between different clusters, as considered in \cite{NewmanZiff00}.  Both of these processes lead to second-order transitions, as do their more dynamic cousins, the directed-percolation and contact processes.  Here we consider the question of whether the PR rule produces a discontinuous transition in the two-dimensional percolation model also.  Indeed, we find that the transition is very sharp and explosive, and find strong indication that the transition is indeed first-order.

We consider bond percolation on $L \times L$ square lattices with periodic boundary conditions in both directions, with $n = L^2$ sites.  Bonds are added randomly and one at a time, keeping track of the current cluster structure of the system (the so-called Newman-Ziff process)\cite{NewmanZiff00}, modified by the consideration of two unoccupied bonds that could connect distinct clusters.  The bond that minimizes the product of the masses of the two clusters it joins is preferentially chosen to become occupied, and the clusters are merged into one.  Time $t$ represents the number of successful bonds added, and each new bond, which always connects different clusters, reduces the number of clusters by one.  Sites with no bonds are considered to be clusters of one site, so initially we have $n$ clusters.  After $t$ bonds are added, the number of clusters $N$ in the system equals $n - t$. \ 
We characterize the mass of a cluster by the number of sites, $s$. \  Because we only add bonds between sites of different clusters, the bonds themselves form ``minimum spanning trees" over each cluster, yet for the conventional growth process, where bonds are added one at a time, standard percolation clusters (characterized by the sites that are connected ) are created.  

\begin{figure}
\includegraphics[scale = 0.25]{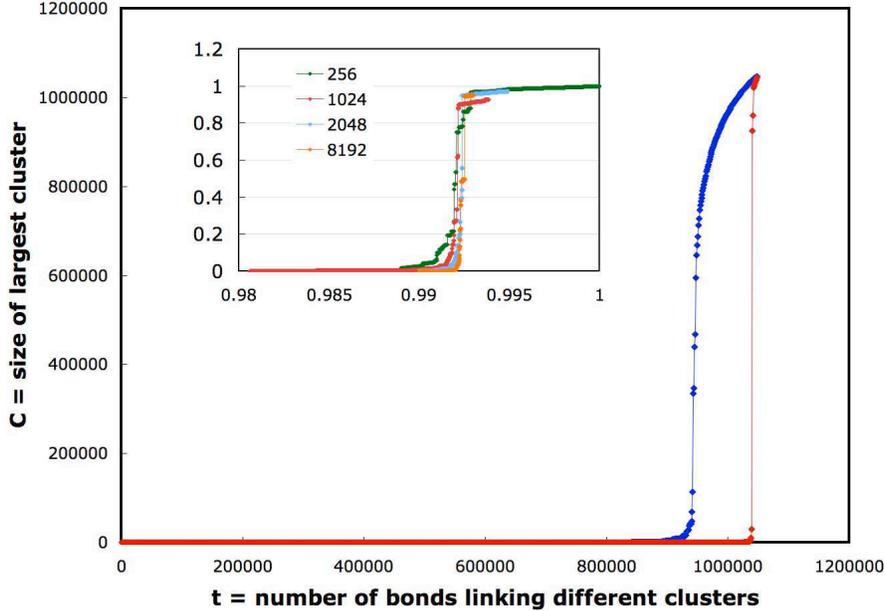}
\caption{\label{fig1} (color online) Plot of regular percolation (blue) and the PR (red) for bond percolation on a lattice of size $1024 \times 1024$, showing the delayed and explosive growth in the PR model.  Points are every $1024$ time steps.  Inset shows PR model on an expanded scale, on systems of sizes $L = 256$, $1024$, $2048$ and $8192$, with the scaled axes $\Delta/n$ vs.\ $t/n$ with $n = L^2$.}
\end{figure}

We have carried out simulations of the cluster growth using both the regular and PR processes, like in Ref.\ \cite {AchlioptasDSouzaSpencer09} measuring the maximum cluster size $C$ as a function of $t$. \ 
The main plot of Fig.\ \ref{fig1} shows the results of these simulations for single realizations on a $1024 \times 1024$ lattice.  In the PR model, the transition is quite sharp.  An expansion of this curve shows that the jump substantially occurs over a small number of time steps, as shown in the inset for different system sizes (here with the axes scaled by $n$).  In contrast, for regular percolation, the transition is much smoother.  The simulations on different size lattices show that the first-order transition in the PR model is robust.

For regular percolation, we can locate the transition point exactly.  While  we don't know the effective overall bond occupation fraction $p$, which includes bonds between sites in a cluster beyond the spanning tree, and thus don't know when the square-lattice threshold $p_c = 1/2$ is reached, we can identify the transition by finding the point where the number of clusters per site reaches its critical value \cite{TemperleyLieb71,ZiffFinchAdamchik97}
\begin{equation}
\frac{N_c}{n} \sim \frac {3 \sqrt{3} - 5}{2} \approx 0.098076 \ ,
\label{Nc}
\end{equation}
which is known to have small finite-size corrections for a periodic system \cite{ZiffFinchAdamchik97}.
This density will be reached when $t/n  = 1 - N_c/n= (7 - 3 \sqrt{3})/2 \approx 0.901924$. \  Of course, for a normal system at $p_c$, there will normally be some fluctuation in $N_c$, but for large systems these fluctuations will be small, so finding where the density exactly equals $1-N_c/n$ gives an excellent estimate of the critical point.  We have verified that, at this point, the average $C$ for different size systems scales as $L^D$ with $D \approx 1.8953$, consistent with the theoretical value $91/48 \approx 1.8958$. \  On the other hand, for the PR rule, we have no {\it a priori} knowledge on where the transition point should be.

\begin{figure}
\includegraphics[scale = 0.5]{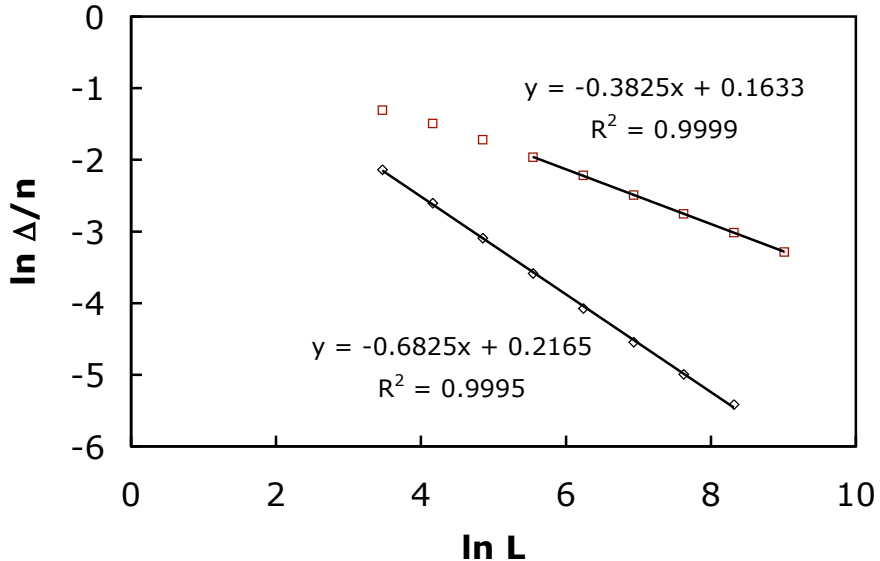}
\caption{\label{fig2} (color online) Scaling of $\Delta/n = (t_0 - t_1)/n$ for regular percolation (upper curve) and the PR model (lower curve) with system size $L$.}
\end{figure}

Achlioptas et al.\ characterized the transitions by two times: the time $t_0$ where $C$ equals $\sqrt{n}$, and the time $t_1$ where $C$ equals $n/2$, and then considered $\Delta = t_1 - t_0$. \   For unbiased growth, they argued that $\Delta$ should be proportional to $n$, while for the PR  growth they found $\Delta \approx n^{2/3}$.  \  We have measured the same quantities for regular the PR growth on square lattices with $L = 32$, $64, \ldots$ $8192$, with the number of realizations ranging from $100000$ for the smaller sizes to $100$ for the largest.  The results for $\Delta$ are plotted in Fig.\ \ref{fig2}.

This figure provides clear evidence that the two transitions are of a quite different nature.  The regular percolation model does not correspond to $\Delta/n$ going to a constant as in the Erd\H os-R\'enyi case, but instead decreases as $\Delta/n \sim L^{-0.383} = n^{-0.192}$.  \  In fact, $t_1$ converges rapidly to $t_c$ (because, for finite lattices of this size, at the critical point $C/n$ is close to $1/2$, which is the condition used to calculate $t_1$).  The main variation in $\Delta$ is due to the variation in $t_0$. \  By normal scaling arguments, we expect the typical cluster mass $s^*$ to scale as $s^* \sim |p - p_c|^{-1/\sigma}$ with $\sigma = 36/91 \approx 0.3956$ in 2d \cite{StaufferAharony94}, so setting $s^* = \sqrt{n} = L$ to correspond to the point $t_0$, and using the fact that $p_0-p_c$ is proportional to $(t_0 - t_c)/n \approx -\Delta/n$ for small $p_0-p_c$, we deduce that
\begin{equation}
\Delta/n \sim n^{-\sigma/2} = L^{-36/91} \ ,
\label{delta}
\end{equation}
which agrees fairly well with our numerical observations.  Actually, the maximum cluster size $C$ differs from the typical cluster size $s^*$ by terms logarithmic in $L$, which would add logarithmic corrections to the above prediction and may account for the small deviations seen in the behavior from (\ref{delta}). \

We emphasize that $p_0-p_c$ is proportional to, but not equal to, $(t_0 - t_c)/n$, because in our growth process, $t = n - N$ reflects the number of bonds linking different clusters added to the system, and does not include bonds within existing clusters.  That is, $p$ is not equal to $t/n$.  The number of clusters $N(p) = n - t$  has a high-order singularity $|p-p_c|^{2-\alpha}$ with $\alpha = - 2/3$ at $p_c$, but this is masked by the lower-order continuous terms.  Thus, it is valid to assume  $(t_0 - t_c)/n \propto p_0-p_c$ to first order.  

For the PR model, we find the much different behavior: $\Delta \sim L^{-0.683}$. \  Closer examination shows that both $t_0-t_c$ and $t_1-t_c$ scale with the same exponent of about $-0.70$, with coefficients $-1.429$ and $0.040$, respectively.  Thus $t_1/n$ changes very little with $L$, and has a well-defined limiting value  $\approx0.9925$ as $L \to \infty$.   This limiting value implies a critical density of clusters $N_c/n \approx 0.0075$ at the transition point, much lower than the value $N_c/n$ for regular percolation given in Eq.\ (\ref{Nc}). 

\begin{figure}
\includegraphics[scale = 0.25]{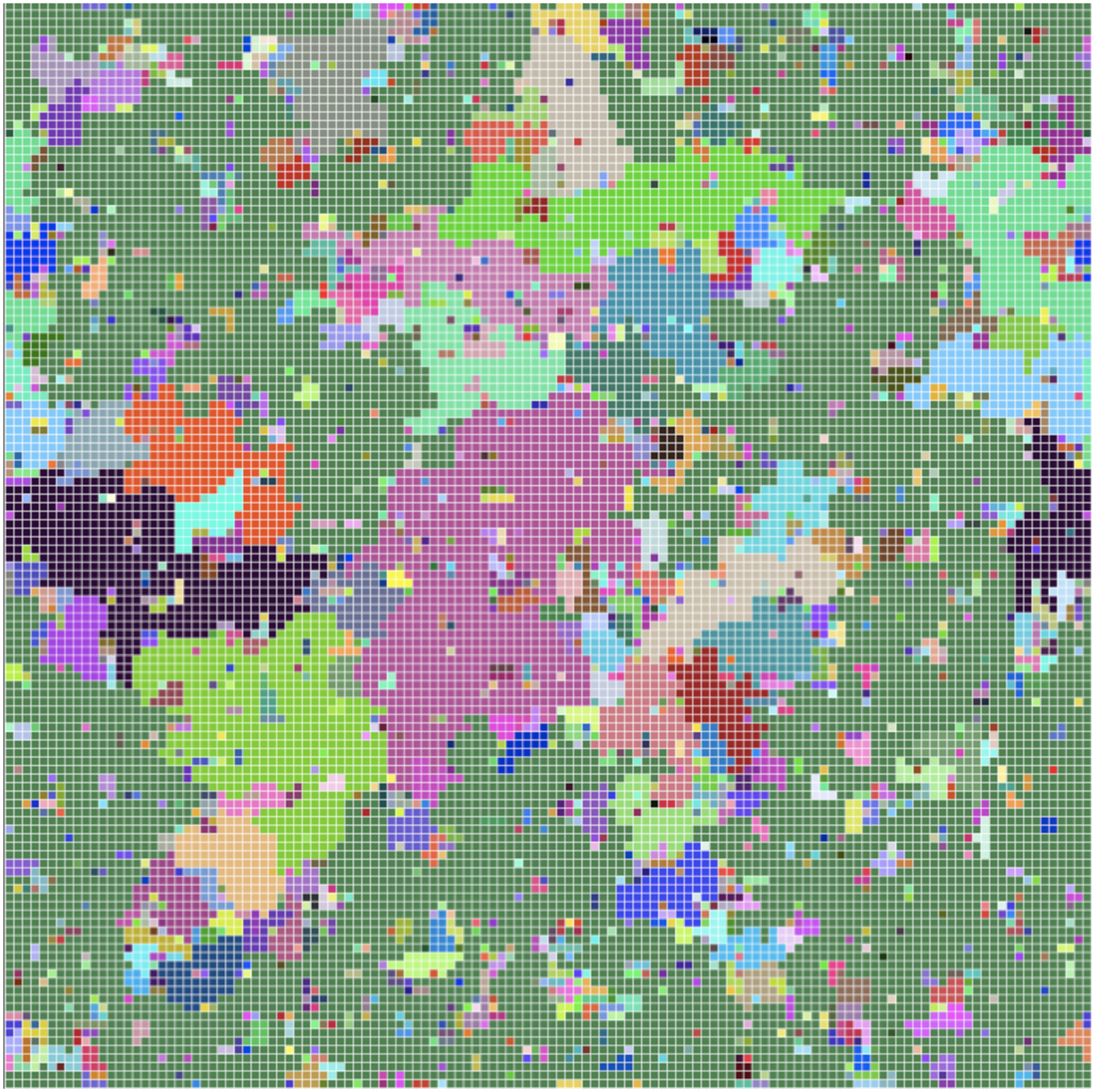}
\includegraphics[scale = 0.25]{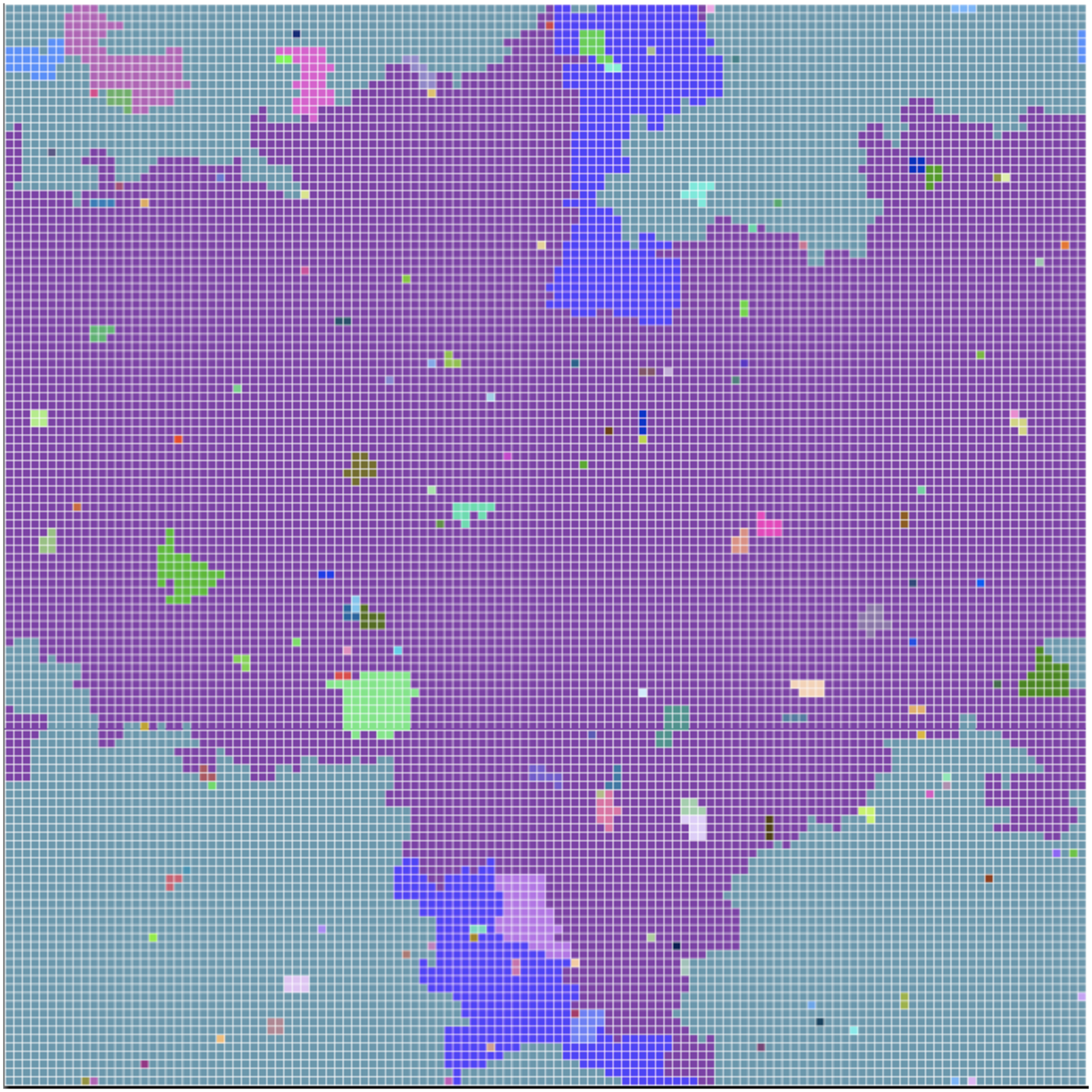}
\caption{\label{fig3} (color online) Regular (left) and PR (right) percolation models at $C \approx n/2$.}
\end{figure}

\begin{figure}
\includegraphics[scale = 0.25]{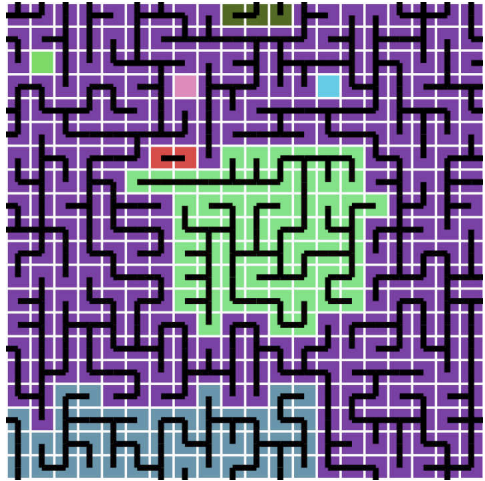}
\caption{\label{fig4} (color online) PR model close-up, showing the spanning-tree bond structure.}
\end{figure}

We illustrate the critical behavior of the two models  in Fig.\ \ref{fig3}, where we show snapshots of the surface for each model for $C \approx n/2$ on a $128 \times 128$ lattice.  Here we show only the sites of the lattices, colored differently according to the clusters they belong to.  The lower number of clusters in the PR model is clearly evident.  We also show in Fig.\ \ref{fig4} a closeup of the PR case, showing the connecting spanning-tree bonds.

This marked difference in the critical behavior between the PR model and regular percolation implies an underlying difference in the nature of the transition.  The behavior in Fig.\ \ref{fig1} points to that transition being of first order in the sense that once the large clusters form, the process rapidly goes to near-completion and the maximum cluster size has a discontinuous nature.  

%The two ``phases" present are the one very large cluster, and the remaining small finite-size clusters in the system.

%In normal cluster growth there is an effective bias toward the joining of larger clusters, which in fact is responsible for the gelation or percolation transition; here we decrease that bias  by choosing the bond with the lower weight.  This has the effect of depleting smaller clusters and then producing explosive growth and a sharper transition.

Having a simple lattice growth model showing strong first-order behavior is useful in modeling  variety of discontinuous and explosive growth processes.  Kinetic growth processes have been the object of extensive study over the last two decades \cite{MarroDickman99}, yet this type of growth behavior is unusual in simple models.   An example of a growth process that leads to behavior almost identical to the PR curves shown in Fig.\ \ref{fig1} is in the dynamics of a catalyst system (oxidation of carbon dioxide) with defects randomly scattered on the catalyst surface \cite{LorenzHaghgooieKennebrewZiff02}.  The defects have the effect of creating regions of different sizes, and as the CO partial pressure increases, regions ``poisoned" with CO coalesce along paths of least resistance and create a very similar explosive growth of surface coverage.   Many questions for the present model, such as fractal properties of the resulting clusters, other critical exponents, dependence upon growth rules, behavior in higher dimensions, etc., remain.

%\begin{acknowledgments}
The author thanks Dietrich Stauffer for encouragement and many helpful comments, and acknowledges support from the U. S. National Science Foundation grant number DMS-0553487.
%\end{acknowledgments}

\end{document}